\documentclass[aps,prsty,floatfix,twocolumn]{revtex4}
\usepackage{graphics}
\usepackage{dcolumn}
\usepackage{epsfig}

\newcommand{\bk}{{\bf k }}

\newcommand{\bq}{{\bf q }}

\begin{document}

\bibliographystyle{prsty}

\title
{Van Hove Singularity and Apparent Anisotropy\\
in the Electron-Phonon Interaction in Graphene}
\author{Cheol-Hwan Park$^{1,2}$}
\email{cheolwhan@civet.berkeley.edu}
\author{Feliciano Giustino$^{1,2}$}
\author{Jessica L. McChesney$^3$}
\author{Aaron Bostwick$^3$}
\author{Taisuke Ohta$^3$}
\author{Eli Rotenberg$^3$}
\author{Marvin L. Cohen$^{1,2}$}
\author{Steven G. Louie$^{1,2}$}
\affiliation{$^1$Department of Physics, University of California at Berkeley,
Berkeley, California 94720 USA\\
$^2$Materials Sciences Division, Lawrence Berkeley National
Laboratory, Berkeley, California 94720 USA\\
$^3$Advanced Light Source, Lawrence Berkeley National Laboratory, Berkeley, California 94720 USA}

\date{\today}

\begin{abstract}
We show that the electron-phonon coupling strength
obtained from the slopes of the electronic energy vs. wavevector dispersion relations,
as often done in analyzing angle-resolved photoemission data,
can differ substantially from the actual electron-phonon coupling strength
due to the curvature of the bare electronic bands.
This effect becomes particularly important
when the Fermi level is close to a van Hove singularity.
By performing {\it ab initio} calculations on doped graphene
we demonstrate that, while the apparent strength
obtained from the slopes of experimental photoemission data
is highly anisotropic, the angular dependence of the actual
electron-phonon coupling strength in this material is negligible.
\end{abstract}
\maketitle

The energies and lifetimes of charge carriers in solids are significantly
affected by many-body interactions including {those with electron-hole pairs},
{plasmons}, and {phonons}.
Angle-resolved photoemission spectroscopy has emerged as an ideal
tool for directly probing the
effects of these interactions on the electron quasiparticle dynamics
with good energy and momentum resolution~\cite{damascelli:2003RMP_ARPES}.

{In particular}, the low-energy electron dynamics at metal
surfaces~\cite{plummer:2003PSurfSci_MetalSurf_ElPh},
in layered materials, such as
magnesium diboride~\cite{souma:2003Nat_MgB2},
graphite~\cite{zhou:2001NatPhys_graphite},
and cuprate superconductors~\cite{damascelli:2003RMP_ARPES},
and in single layer
graphene~\cite{bostwick:2007NatPhys,ohta:2007PRL_Graphene_ARPES,bostwick:2007NJP,
mcchesney:2007arxiv_Graphene_Anisotropy,ohta:2008NJP,mucha-kruczynski:2007arXiv,
zhou:2007NatMat,zhou:2008arXiv,
park:2007PRL_Graphene_ElPh,
calandra:2007PRB_Graphene_Dopant,calandra:2007PRB_Graphene_ElPh,tse:2007PRL_Graphene_ElPh},
is significantly affected by the electron-phonon {interaction}.
Since the electron-phonon interaction generally manifests itself as
a kink in the quasiparticle dispersion
relations measured by angle-resolved photoemission
spectroscopy~\cite{engelsberg:1963PR_ElPh,grimvall:1981_Metal_ElPh},
it is common practice to determine the strength of the electron-phonon coupling
by taking the ratio between the group velocity at the Fermi level
and below the phonon-induced
kink~\cite{hengsberger:1999PRL_ElPh,hengsberger:1999PRB_ElPh,
rotenberg:2000PRL_ElPh,mcchesney:2007arxiv_Graphene_Anisotropy}.
In the cases where this simple procedure is not applicable,
more complicated self-consistent
algorithms~\cite{kaminski:2005NJP_ARPES,kordyuk:2005PRB_ARPES} become necessary
to extract the electron-phonon coupling strength. However, the application of these
methods requires the knowledge of several adjustable parameters and is subject to some
arbitrariness. Therefore, assessing in the first instance the validity of
extraction procedures based on
the linear slopes of the photoemission data is an important issue.

Graphene~\cite{novoselov:2004Sci_Graphene,
novoselov:2005PNAS_2D,novoselov:2005Nat_Graphene_QHE,
zhang:2005Nat_Graphene_QHE,berger:2006Sci_Graphene_Epitaxial}
is an ideal system to investigate these effects.
Indeed, the Fermi level of graphene can be tuned over a wide
energy range by chemical
doping~\cite{bostwick:2007NatPhys_Graphene,mcchesney:2007arxiv_Graphene_Anisotropy}
or by gating~\cite{novoselov:2005Nat_Graphene_QHE,zhang:2005Nat_Graphene_QHE,
geim:2007NatMat_Graphene_Review},
and can almost be aligned with the van Hove singularity
at the M point of the two-dimensional Brillouin
zone~\cite{mcchesney:2007arxiv_Graphene_Anisotropy}.

In this work, we show that
the {apparent} electron-phonon coupling strength in doped graphene
obtained from the linear slopes of the
renormalized quasiparticle dispersions, as calculated from first principles,
is highly {\it anisotropic}, in good agreement with experimental
results~\cite{mcchesney:2007arxiv_Graphene_Anisotropy}.
On the other hand,
the phonon-induced electron self-energy
is found to be only weakly dependent on the wavevector in the
Brillouin zone. As a consequence,
the actual electron-phonon coupling strength is
{\it isotropic}.
The apparent anisotropy of the electron-phonon interaction
is shown to arise from the
curvature of the bare electronic bands of graphene, {which is}
strongly enhanced in proximity of the van Hove singularity at the M point.
{Our findings are relevant} to the interpretation
of photoemission spectra in materials
where the Fermi level is aligned with a van Hove singularity,
such as {the} hole-doped cuprates at optimal doping.

The mass-enhancement parameter or electron-phonon coupling strength
$\lambda_\bk$ of an electronic state with wavevector
$\bk$ on the Fermi surface can be expressed through the energy
derivative of the real part of the self-energy arising from
the electron-phonon interaction~\cite{nakajima:1963PTP_Metal_ElPh}:
\begin{equation}
\lambda_\bk =-\left.\frac{\partial\ {\rm Re}\Sigma_\bk(E)}
{\partial E}\right|_{E=E_{\rm F}}\ ,
\label{equation:lambda}
\end{equation}
$E_{\rm F}$ being the Fermi level.
Within the Migdal approximation, which corresponds to considering
the non-crossing electron-phonon self-energy diagrams,
this quantity can be calculated
by~\cite{grimvall:1981_Metal_ElPh, giustino:2007TBP_Migdal}:
  \begin{eqnarray}
  &&\lambda_\bk
 = \sum_{m,\nu} \, \int \, \frac{d\bq}{A_{\rm BZ}}
  \; |g_{mn,\nu}(\bk,\bq)|^2 \nonumber \\
  &&\hspace{-0.cm}\times\ \left[\frac{n_{\bq\nu}
  +1-f_{m\bk+\bq}}
  {(E_{\rm F}-\epsilon_{m\bk+\bq}-\omega_{\bq\nu})^2}
  + \frac{n_{\bq\nu}+f_{m\bk+\bq}}
  {(E_{\rm F}-\epsilon_{m\bk+\bq}+\omega_{\bq\nu})^2}
  \right]\ , \nonumber \\ 
  \label{equation:lambda_abinitio}
  \end{eqnarray}
where $\epsilon_{n\bk}$ is the energy of an electron in the band $n$
with wavevector $\bk$, $\omega_{\bq\nu}$ the energy of a phonon
in the branch $\nu$ with wavevector $\bq$, and
$f_{n\bk}$ and $n_{\bq\nu}$ are the Fermi-Dirac and Bose-Einstein
occupations, respectively.
The integration is performed within the two-dimensional Brillouin zone of area $A_{\rm BZ}$.
The electron-phonon matrix element
$g_{mn,\nu}(\bk,\bq)=\langle m\bk+\bq|\Delta V_{\bq\nu}|n\bk \rangle$
is the amplitude for the transition from an electronic state
$|n\bk \rangle$ to another state $|m\bk+\bq \rangle$ {induced by}
the change in the self-consistent potential
$\Delta V_{\bq\nu}$ generated by the phonon $\left|\bq\nu\right>$.
The technical details of the calculations are reported
in Ref.~\onlinecite{park:2007PRL_Graphene_ElPh}.

It can be shown~\cite{grimvall:1981_Metal_ElPh}
that the actual electron-phonon coupling strength in Eq.~(\ref{equation:lambda})
can also be written as
\begin{equation}
\lambda_\bk=
\frac{v^0_\bk(E_{\rm F})}{v_\bk(E_{\rm F})}-1\ ,
\label{equation:lambda_true}
\end{equation}
where $v_\bk(E_{\rm F})$ and $v^0_\bk(E_{\rm F})$
are the renormalized and the bare velocities at the Fermi level, respectively.
While Eq.~(\ref{equation:lambda_true}) is useful for theoretical analyses,
it cannot be used directly
in determining the electron-phonon coupling strength from the experimental data
since the bare group velocity is merely a conceptual tool and cannot be measured.
To circumvent this difficulty, from the experimentally measured low-energy photoemission spectrum,
the electron-phonon coupling strength is usually
extracted~\cite{damascelli:2003RMP_ARPES,mcchesney:2007arxiv_Graphene_Anisotropy}
by taking the ratio of the renormalized velocity
below and above the phonon kink.
This procedure rests on the assumptions that (i) well beyond the phonon energy scale
the bare velocity is fully recovered and (ii) the bare band is linear over the
energy range considered.
We denote the value obtained from this procedure as the apparent electron-phonon
coupling strength:
\begin{equation}
\lambda^{\rm app}_\bk=
\frac{v_\bk(E_{\rm F}-\Delta E)}{v_\bk(E_{\rm F})}-1\ ,
\label{equation:lambda_app}
\end{equation}
where $\Delta E$ is taken slightly larger than the phonon energy $\omega_{\rm ph}$
so that the energy $E= E_{\rm F}-\Delta E$ falls below the
phonon kink.

  \begin{figure*}
  \includegraphics[width=2.0\columnwidth]
  {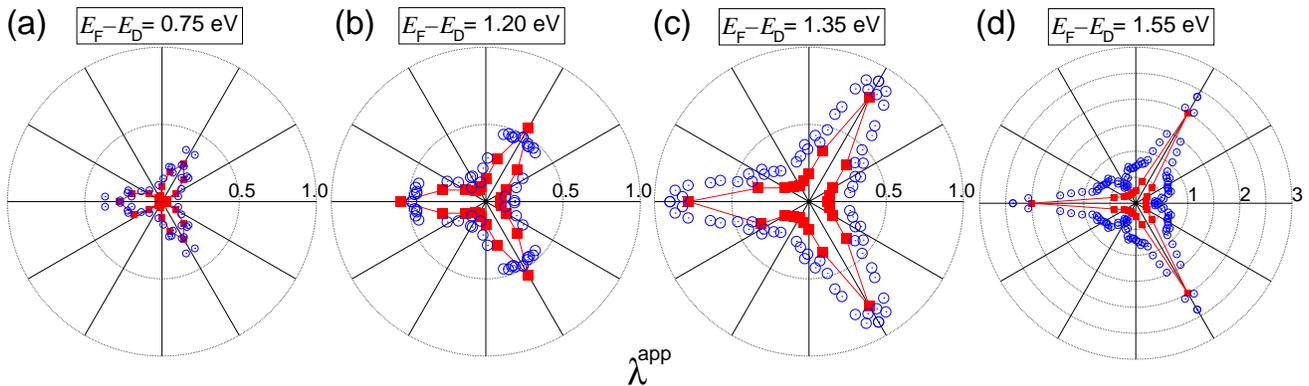}
  \caption{(color online) Polar plots of the apparent electron-phonon coupling strength
  $\lambda^{\rm app}_\bk$ {on the Fermi surface} around the K point
  in the Brillouin zone.
  Filled red squares and empty blue circles
  represent results from the {\it ab initio}
  calculation and from the experimental photoemission spectra, respectively.
  The lines are included as a guide to the eye.
  Different panels correspond to different doping levels.
  Note that the scale in (d) is different from that in (a)-(c).}
  \label{Figure_Anisotropy_4panels}
  \end{figure*}

In Fig.~\ref{Figure_Anisotropy_4panels} we compare the
apparent coupling strength $\lambda^{\rm app}_\bk$
obtained from
our first-principles calculations and that extracted from
the experimental photoemission spectra of graphene
at four different levels of doping~\cite{mcchesney:2007arxiv_Graphene_Anisotropy}.
To determine $v_\bk(E_{\rm F}-\Delta E)$
in Eq.~(\ref{equation:lambda_app}) from our first-principles calculations,
we considered the slope of the
quasiparticle band at the energy $\Delta E=$0.3~eV below the Fermi
level~\cite{note:DeltaE}.
We have checked that as the energy range $\Delta E$
varies within the interval 0.2$\sim$0.4~eV,
the apparent strength $\lambda^{\rm app}$ changes by less than 10~\%
along both the KM and K$\Gamma$ directions.
Theory and experiment are in
good agreement with each other for all doping levels considered.
For graphene at the highest doping level ($E_{\rm F}-E_{\rm D}=1.55$~eV,
{where} $E_{\rm D}$ {being the energy at the Dirac point})
the agreement between theory and experiment is slightly worse along the
K$\Gamma$ direction.

\begin{figure}
\includegraphics[width=1.0\columnwidth]{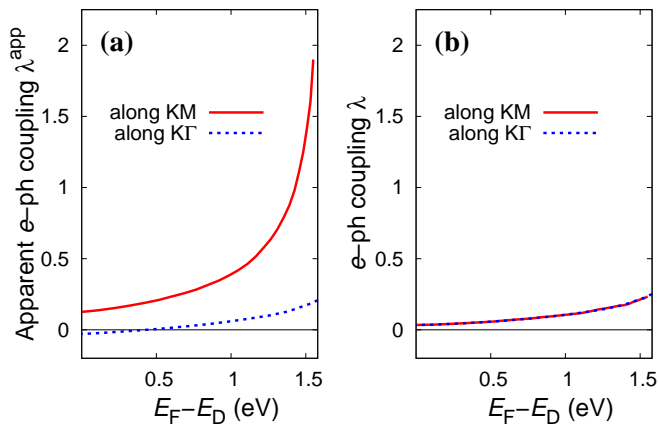}
\caption{(color online)
  The apparent strength $\lambda^{\rm app}$ (a) and
  the actual electron-phonon coupling strength $\lambda$ (b)
  calculated along two different directions in the two-dimensional
  Brillouin zone of graphene: along KM (solid red lines)
  and along K$\Gamma$ (dashed blue lines). {Along the K$\Gamma$} direction,
  $\lambda^{\rm app}$ can even become negative {[cf. discussion around
  Eq.~(\ref{equation:lambda_app_approx})]}.}
\label{Figure_Lambda}
\end{figure}

Figure~\ref{Figure_Lambda}(a) shows the calculated apparent strength
$\lambda^{\rm app}$ [Eq.~(\ref{equation:lambda_app})]
as a function of doping for two
different directions in the Brillouin zone of graphene.
The apparent electron-phonon coupling strength
is highly anisotropic and can become as large as $\lambda^{\rm app}=2$
along the KM direction
for the doping levels considered here.

Now we consider the actual electron-phonon coupling strength $\lambda$ as obtained from
Eq.~(\ref{equation:lambda_abinitio}) [Fig.~\ref{Figure_Lambda}(b)].
The actual strength  increases monotonically
with doping, reaching $\lambda=0.22$ when $E_{\rm F}-E_{\rm D}=1.5$~eV.
At variance with the apparent strength, the actual strength $\lambda$ does not depend on
the direction of the wavevector $\bk$.
We have checked that this holds
for any path through the K point.
The present results indicate that the actual electron-phonon coupling strength in
doped graphene is extremely isotropic.
Thus, the actual strength
can differ substantially
from the apparent strength,
the more so as the Fermi surface approaches
the van Hove singularity at the M point.

\begin{figure}
\includegraphics[width=1.0\columnwidth]{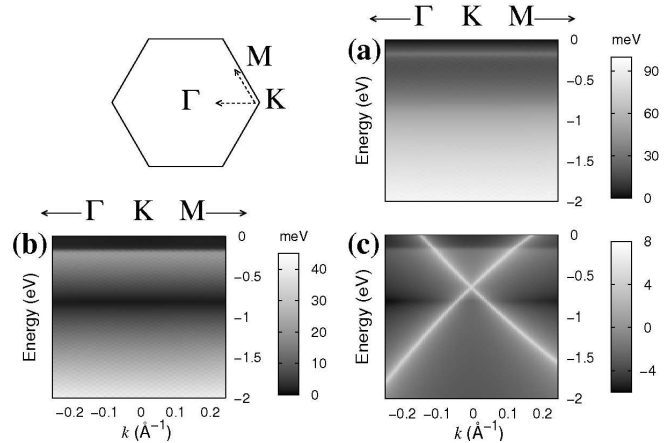}
\caption{Calculated (a) real and (b) imaginary part of the electron self-energy $\Sigma(E,\bk)$
  and (c) logarithm of the corresponding spectral function $A(E,\bk)$ arising from the
  electron-phonon interaction in $n$-doped graphene ($E_{\rm F}-E_{\rm D}=0.64$~eV),
  along two different directions KM and K$\Gamma$ in the Brillouin zone.}
\label{Figure_Selfenergy}
\end{figure}

In order to analyze in detail the angular dependence of the
electron-phonon coupling in graphene,
we calculated the phonon-induced electron self-energy
using Eq.~(1) of Ref.~\onlinecite{park:2007PRL_Graphene_ElPh}. While in
Ref.~\onlinecite{park:2007PRL_Graphene_ElPh}
the electron self-energy was evaluated under the constraint $E=\epsilon_\bk$,
we here consider the complete energy-dependent self-energy $\Sigma_\bk(E)$.
Figure~\ref{Figure_Selfenergy} (a) and (b) show the real and the
imaginary part of the electron self-energy in
n-doped graphene ($E_{\rm F}-E_{\rm D}=0.64$~eV).
The wavevector is varied along two different paths
indicated in the upper left corner of Fig.~\ref{Figure_Selfenergy}.
The dependence of the self-energy
on the wavevector $\bk$ is found to be extremely weak,
the variation along the path considered in Fig.~\ref{Figure_Selfenergy}
being less than 3~meV for a given energy $E$.
The insensitivity of the electron-phonon coupling strength $\lambda_\bk$
to the wavevector $\bk$ [see Fig.~\ref{Figure_Lambda}(b)]
is fully consistent with the finding on the self-energy.
Figure~\ref{Figure_Selfenergy} also shows that, while the self-energy is highly isotropic,
the corresponding spectral function exhibits significant angular dependence due to
the anisotropic dispersion of the energy bands in graphene.

Recently, the observed anisotropy~\cite{mcchesney:2007arxiv_Graphene_Anisotropy}
in the apparent electron-phonon coupling strength has been related to foreign atoms
based on calculations of CaC$_6$ layers with the dopants arranged periodically
in atop sites on the graphene plane~\cite{calandra:2007PRB_Graphene_Dopant}.
Our calculations clearly show that the anisotropy in the apparent strength
$\lambda^{\rm app}$ is already present without invoking the possible effect of the dopants.

\begin{figure}
\includegraphics[width=1.0\columnwidth]{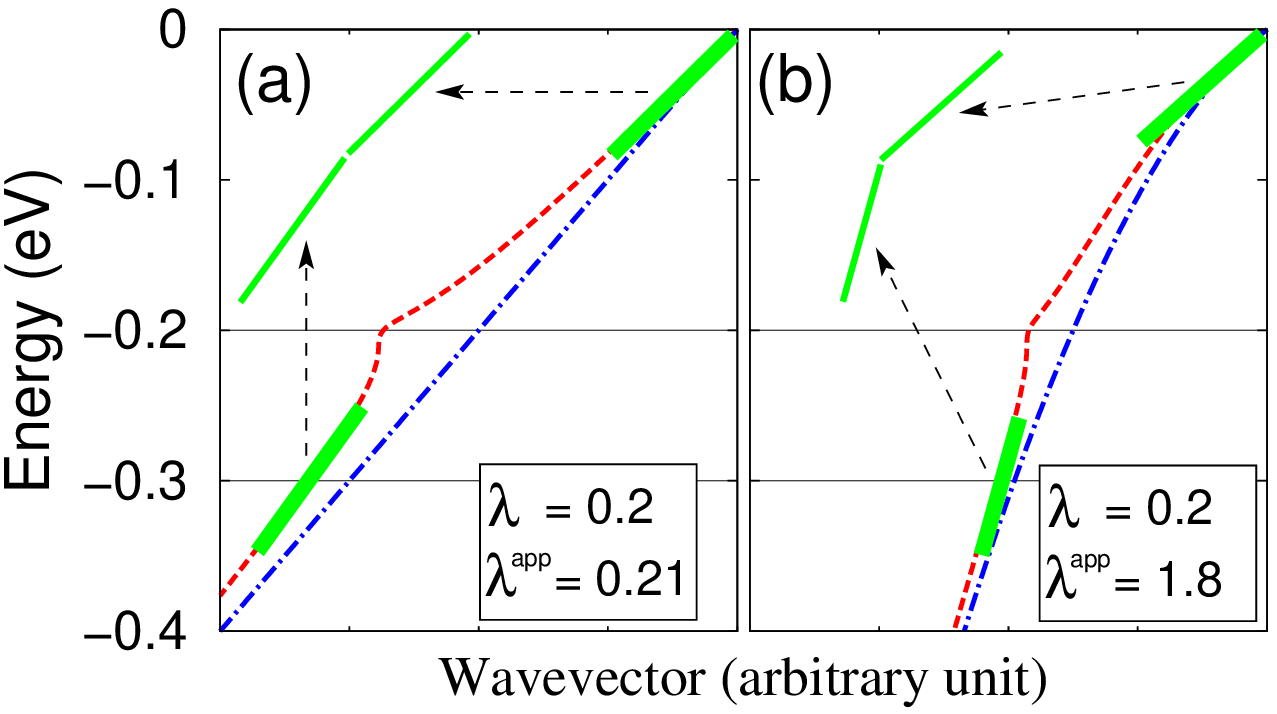}
\caption{(color online) Quasiparticle bandstructures of model systems
  including the electron-phonon interaction (dashed red lines).
  In (a) and (b) the bare electronic
  bands (dash-dotted blue lines) are {assumed to be} linear and quadratic, respectively.
  In each case, the actual electron-phonon coupling strength is set to $\lambda=0.2$.
  The horizontal solid lines represent the phonon energy $\omega_{\rm ph}=0.2$~eV, and
  the energy $\Delta E=0.3$~eV below the Fermi level ($E_{\rm F}=0$)
  at which the slope is taken to calculate the apparent strength $\lambda^{\rm app}$.
  The solid green line segments are tangential to the quasiparticle bandstructure at
  $E=0$ or $E=-\Delta E$.}
\label{Figure_Schematic}
\end{figure}

Our investigation of doped graphene allows us to discuss some general
aspects of the extraction of the electron-phonon coupling parameters
from angle-resolved photoemission data.
By expanding the energy dependence of the velocity to first order, we can rewrite
{approximately}
the apparent electron-phonon coupling strength in Eq.~(\ref{equation:lambda_app}) as:
\begin{equation}
\lambda^{\rm app}_\bk \approx \lambda_\bk-\frac{v'^0_\bk(E_{\rm F})}{v_\bk(E_{\rm F})}\Delta E\ .
\label{equation:lambda_app_approx}
\end{equation}
In Eq.~(\ref{equation:lambda_app_approx}), $v'^0_\bk(E_{\rm F})$ is the energy derivative
of the bare velocity, and we assumed that well below the phonon kink
the bare and the renormalized velocities coincide.
Equation~(\ref{equation:lambda_app_approx}) shows that,
whenever the band velocity decreases with decreasing binding energy {(i.e., $v'^0_\bk<0$)}
as is the case for graphene along the KM direction,
the apparent electron-phonon coupling strength always exceeds the actual strength.

To illustrate this point, we consider in Fig.~\ref{Figure_Schematic} the quasiparticle bandstructure
for a model system obtained by assuming an Einstein phonon spectrum with phonon
energy $\omega_{\rm ph}=0.2$~eV, a constant density of states near the Fermi level,
and a constant electron-phonon coupling strength $\lambda$=0.2.
Within this model, the real part of the electron self-energy due to electron-phonon interaction
reads~\cite{engelsberg:1963PR_ElPh}:
\begin{equation}
{\rm Re}\ \Sigma(E)=-\frac{\lambda\omega_{\rm ph}}{4}{\rm log}\left|
\frac{(E+\omega_{\rm ph})^2+\Gamma^2}{(E-\omega_{\rm ph})^2+\Gamma^2}\right|,
\label{equation:simplemodel}
\end{equation}
having included broadening $\Gamma=10$~meV for convenience.
As shown in Fig.~\ref{Figure_Schematic}(a), the apparent strength $\lambda^{\rm app}=0.21$
constitutes a good approximation to
the actual strength $\lambda=0.2$ when the slope of the bare electronic band does not
change appreciably within the phonon energy scale, i.e.
$v'^0_\bk(E_{\rm F})\Delta E\ll v_\bk(E_{\rm F})$.
However, in the case where the bare velocity decreases with decreasing binding
energy, the apparent strength $\lambda^{\rm app}=1.8$ differs significantly from
the actual electron-phonon coupling strength $\lambda=0.2$,
consistent with Eq.~(\ref{equation:lambda_app_approx}) [Fig.~\ref{Figure_Schematic}(b)].
In the limiting situation where
the Fermi level is aligned with the van Hove singularity
(as in heavily doped graphene),
the velocity at the Fermi level vanishes while
the velocity below the phonon kink energy
remains finite. As a result, the apparent strength
obtained through Eq.~(\ref{equation:lambda_app}) becomes exceedingly large.

In conclusion, we have shown that while the phonon-induced electronic self-energy of graphene
is isotropic and consequently the angular dependence of
the electron-phonon coupling strength is negligible,
the apparent electron-phonon coupling strength extracted from
the experimental angle-resolved photoemission spectra using Eq.~(\ref{equation:lambda_app})
exhibits significant anisotropy due to the curvature of the underlying bare electronic bands.
Our analysis indicates that the band curvature
is a crucial ingredient  for the interpretation of angle-resolved photoemission spectra.
For example, the present result may carry implications in the interpretation of
many-body renormalization effects along the antinodal
cuts in the photoemission spectra of cuprate superconductors,
due to the presence of a saddle-point van Hove singularity along the Cu-O bond
directions~\cite{Kordyuk:2002PRL_cuprate,gromko:2002arxiv_cuprate}.

This work was supported by NSF Grant
No. DMR07-05941 and by the Director, Office of Science, Office of Basic Energy
Sciences, Division of Materials Sciences and Engineering Division,
U.S. Department of Energy under Contract No. DE- AC02-05CH11231.
Computational resources have been provided by
the National Partnership for Advanced Computational Infrastructure
(NPACI) and
the National Energy Research Scientific Computing Center (NERSC).
The calculations were performed using the
{\tt Quantum-Espresso}~\cite{baroni:2006_Espresso}
and {\tt Wannier}~\cite{mostofi:2006_Wannier} packages.


\begin{thebibliography}{10}
\bibitem{note:DeltaE}
This specific value of $\Delta E=0.3$~eV was chosen under a reasonable
  assumption that $v_\bk(E_{\rm F}-\Delta E)$ appearing in Eq.~(3) is obtained
  as the average slope between $E=E_{\rm F}-\omega_{\rm ph}$ and $E=E_{\rm
  F}-2\omega_{\rm ph}$. We have checked that as the energy range $\Delta E$
  varies within the interval 0.2$\sim$0.4~eV, the apprent strength
  $\lambda^{\rm app}$ changes less than 10~\% along both KM and K$\Gamma$
  directions.
\end{thebibliography}
\end{document}